\documentclass[10pt,final]{article}
\usepackage{graphicx}
\usepackage{amssymb}
\usepackage{amsmath}
\usepackage{bm}
\usepackage{footmisc}

\usepackage[superscript,biblabel]{cite}
\usepackage{multicol,caption}
\newenvironment{Figure}
{\par\medskip\noindent\minipage{\linewidth}}
{\endminipage\par\medskip}

\begin{document}

\vspace*{1cm}

{\sffamily\huge\bfseries\noindent Quasi-phase-matched nonlinear optical frequency conversion in on-chip whispering galleries}

\vspace{0.5cm} \renewcommand{\thefootnote}{\fnsymbol{footnote}}
{\sffamily\small\noindent Richard Wolf$^{1,3,}$\footnote[1]{These authors contributed equally to this work.},\renewcommand*{\thefootnote}{\arabic{footnote}} Yuechen Jia\footnote{Department of Microsystems Engineering - IMTEK, University of Freiburg, Georges-K\"{o}hler-Allee 102, 79110 Freiburg, Germany.}$^{,*}$, 
Sebastian Bonaus$^{1}$, Christoph S. Werner$^{1}$, Simon J. Herr$^{1}$,
Ingo Breunig$^{1,}$\footnote{Fraunhofer Institute for Physical Measurement Techniques IPM, Heidenhofstra\ss e 8, 79110 Freiburg, Germany.}, Karsten Buse$^{1,2}$ and Hans Zappe\footnote{Gisela and Erwin Sick Chair of Micro-Optics, Department of Microsystems Engineering - IMTEK, University of Freiburg, Georges-K{\"o}hler-Allee 102, 79110 Freiburg, Germany.}

\vspace*{0.3cm}

\textbf{Chip-integrated whispering-gallery resonators enable compact and wavelength-agile nonlinear optical frequency synthesizers. So far, the most flexible phase-matching technique, i.e. quasi phase matching, has not been applied in this configuration. The reason is the lack of suitable thin films with alternating crystal structure on a low-refractive-index substrate. Here, we demonstrate an innovative method of realizing thin film substrates suitable for quasi phase matching by field-assisted domain engineering of lithium niobate, and subsequent direct bonding and polishing. We are able to fabricate high-$\mathbf{Q}$ on-chip WGRs with these substrates by using standard semiconductor manufacturing techniques. The $\mathbf{Q}$-factors of the resonators are up to one million, which allows us to demonstrate quasi-phase-matched second-harmonic generation in on-chip WGRs for the first time. The normalized conversion efficiency is $\mathbf{9\times 10^{-4}\,\textrm{mW}^{-1}}$. This method can also be transferred to other material systems.}
\vspace*{0.3cm}
\begin{multicols}{2}

Whispering-gallery resonators (WGRs) have proven to be highly attractive for nonlinear optics due to their high $Q$-factors, small mode volumes and mechanical robustness \cite{Ilchenko06, Breunig16, Strekalov16}. The achievable high internal intensities lead to efficient frequency conversion already at low input powers in the $\mu$W-region \cite{Furst10}. Made of non-centrosymmetric materials, they are applied for optical-parametric oscillators working from the visible \cite{Werner12} to the mid-infrared \cite{Meisenheimer17} spectral region. They are used for frequency doublers generating light from the ultraviolet \cite{Furst15} to the near infrared as well as for infrared upconversion detectors \cite{Strekalov14, Rueda16b}. To achieve phase matching, methods like birefringent- \cite{Furst10b, Lin15, Wang14b}, cyclic- \cite{Lin13, Lin16} and quasi phase matching \cite{Beckmann11} have been employed. The latter gives ultimate flexibility regarding wavelengths and polarizations of the interacting light fields. Furthermore, it always provides  access to the largest second-order nonlinear-optical coefficient of the material, e.g. in lithium niobate (LiNbO$_3$, LN) this is $d_{333}=27\,\mathrm{pm/V}$ compared with $d_{311}=5\,\mathrm{pm/V}$ \cite{Nikogosyan05}, which has to be used for birefringent phase matching.

The majority of WGR frequency converters are based on bulk resonators. However, recently on-chip WGRs, e.g. out of lithium niobate \cite{Guarino07,Luo17,Wang17b}, aluminum nitride \cite{Guo16}, aluminum gallium arsenide \cite{Mariani14} and gallium nitride \cite{Xiong11} have become very attractive. This is because of their small dimensions in the micrometer region and thus extremely small mode volumes, and the additional potential of building mixed photonic and electrical circuits \cite{Wang17c}, as well as the prospect of using parallel and reproducible semiconductor manufacturing techniques \cite{Wolf17}. On-chip WGRs are fabricated from thin-film substrates to achieve a high refractive index step for sufficient light confinement. They have currently one major limitation: Quasi phase matching is hard to achieve. Although, in general, in ferroelectric crystals the domains can be patterned with the help of external electric fields, field-assisted domain inversion of thin films on substrates is difficult, since the backside of the thin film cannot be accessed electrically. For non-ferroelectric materials the situation is also difficult as one needs to grow periodically structured thin films, which cannot be done on any substrate.

We overcome this limitation for on-chip frequency converters here by introducing a method to obtain thin films with patterns of inverted second-order optical nonlinearity starting with patterned bulk material, which is bonded onto a substrate with lower refractive index and polished down to a thin film. This approach works for many material combinations. In this contribution, we demonstrate this method for periodically-poled lithium niobate on quartz (pp-LNoQ).

The fabrication starts with field-assisted domain inversion of bulk LN by writing calligraphically a linear grating with a period length of $23~\mu\textrm{m}$ into a  300-$\mu\textrm{m}$-thick 5-mol.$\%$-MgO-doped optically-polished $z$-cut-LN chip (Fig.~\ref{fig:Herstellung}). A chromium layer serves as the backside electrode on the +$z$ side of the crystal and we write the domains with the help of a tungsten-carbide tip on the -$z$ side of the wafer, moving along the $y$-crystal axis. A detailed description of the poling procedure can be found in \cite{Werner17}. Next, we clean the periodically poled sample and also a $z$-cut-$\alpha$-quartz chip, followed by a further cleaning and surface- functionalization procedure (for details see supplementary part). The aim of this procedure is to assure perfectly clean and OH-terminated surfaces which are crucial for the direct bonding process \cite{Plol99, Haisma94}.

In a next step, we put the periodically poled sample and the quartz chip together. Hydrogen bonds at the OH-terminated surface already form at room temperature such that the samples stick together. Hereafter, we increase the bond forces by tempering on a hot plate at $325\,^\circ\textrm{C}$ for 5 hours. We choose $\alpha$ quartz as the substrate since the thermal-expansion coefficients $\alpha$ parallel to the a-crystal axis is close to that of LN ($\alpha_{a,\,\mathrm{quartz}}=12.4\times 10^{-6}\,\mathrm{K}^{-1}; \alpha_{a,\,\mathrm{LN}}=14.8\times 10^{-6}\,\mathrm{K}^{-1}$\cite{Weber03}). Subsequently, we reduce the thickness of the periodically poled LN to $2~\mu\textrm{m}$ by lapping and polishing with a standard wafer-polishing machine (\textit{PM5, Logitech}, details see supplementary part).            

\begin{Figure}[h]
	\centering
	\includegraphics[width=8.4 cm]{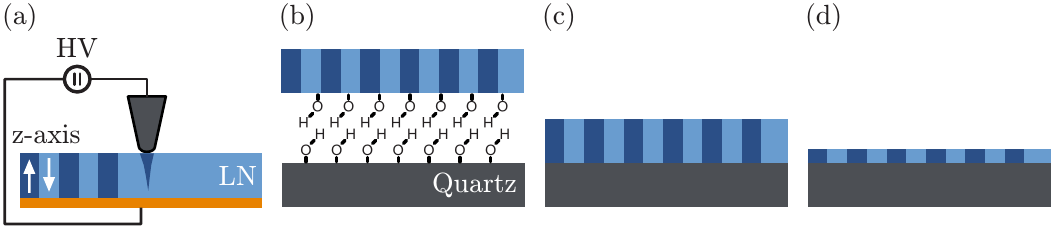}
	\captionof{figure}{Process flow for the fabrication of periodically-poled-lithium-niobate-on-quartz (pp-LNoQ) substrates. (a) Domain inversion; (b) Cleaning, surface functionalization; (c) Thermal bonding; (d) Lapping and polishing.}
	\label{fig:Herstellung}
\end{Figure}

To obtain on-chip WGRs, we structure micro-rings with a diameter of $216~\mu\mathrm{m}$ into the pp-LNoQ thin films by standard semiconductor manufacturing techniques, including lithography, reactive-ion etching and a polishing process. A detailed description can be found in \cite{Wolf17}. Figure~\ref{fig:SEM-WGR} shows the final result. For this image, we made the domains visible by etching a dummy sample in a 40-$\%$-KOH solution at $90\,^\circ\mathrm{C}$ for 45 minutes, since the -$z$ surface of LN etches faster than the +$z$ surface. The linear domain structure with a $23\,\mu \mathrm{m}$ periodicity and a $5\,\mu \mathrm{m}$ domain width can be clearly seen.

\begin{Figure}[h]
	\centering
	\includegraphics[width=8.4 cm]{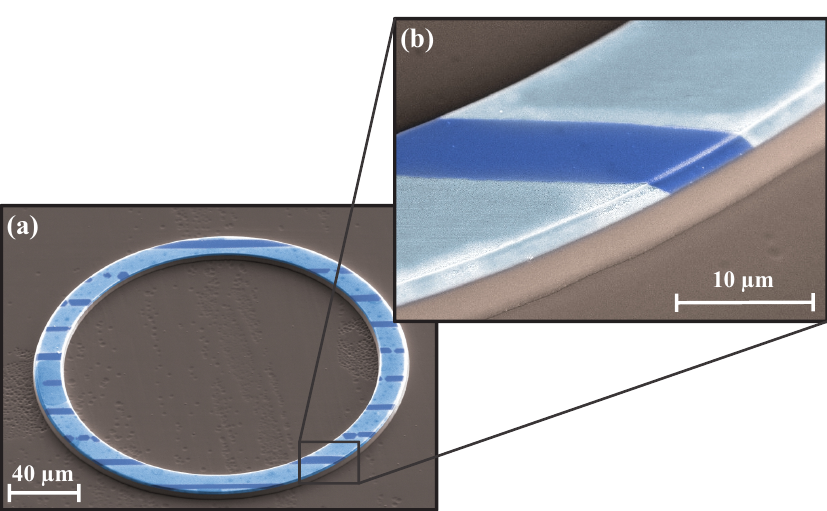}
	\captionof{figure}{False-color SEM images of (a) a pp-LNoQ on-chip WGR with a diameter of $216~\mu\mathrm{m}$ and (b) zoomed view. Dark blue: LN with $z$-axis pointing up. Light blue: LN with $z$-axis pointing down. Brown: quartz-substrate.}
	\label{fig:SEM-WGR}
\end{Figure}

For optical characterization we use the setup shown in Fig.~\ref{fig:Messaufbau}. A DFB laser diode, which emits at 1551 nm and is tunable over 2 nm, serves as the light source. To couple light into the WGR, we use a setup with a fixed chip containing a straight coupling waveguide and a second chip on top comprising the WGRs. We couple light into the coupling waveguide via end-fire coupling with a 40x microscope objective lens. Light penetrates from the coupling waveguide to the WGR via evanescent field coupling. Both the horizontal position of the WGR with respect to that of the coupling waveguide and the vertical distance between the coupling waveguide and the WGR can be adjusted via a $x,y,z$-piezo-actuator stage and hence the coupling strength can be tuned. 

\begin{Figure}[h]
	\centering
	\includegraphics[width=8.4 cm]{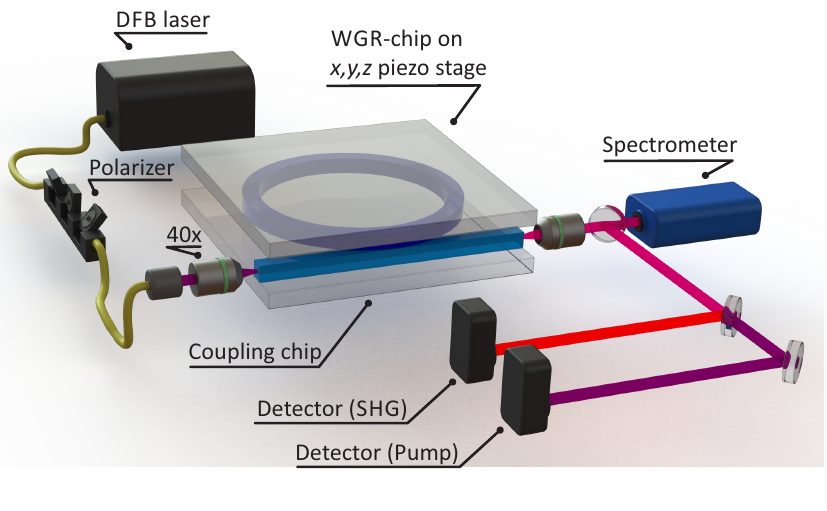}
	\captionof{figure}{Schematic of the setup for $Q$-factor measurement and investigation of second-harmonic generation}
	\label{fig:Messaufbau}
\end{Figure}

First, we determine the $Q$-factor of ordinarily (o) and extraordinarily (e) polarized whispering-gallery modes (WGMs). The first is in the range of $4\times10^5$ to $1.3\times10^6$ while the second is significantly lower, ranging from $7\times10^4$ to $2\times10^5$. We are able to reach critical coupling and overcoupling for o-polarized light only. The reason for the higher loss for e-polarized modes  most likely stems from the fabrication used for the on-chip WGRs: After structuring of ring resonators into the top LN thin film by lithography and reactive-ion etching, we use a polishing process to reduce the side-wall roughness and thus the scattering loss \cite{Wolf17}. To clean the chips after polishing and to remove the polishing particles, we deploy a wet-chemical cleaning procedure with KOH. This is uncritical and does not affect the quality factor of either the o- nor the e-polarized modes for resonators having no domain pattern \cite{Wolf17}. However, since KOH etches the -z surface of LN faster than the +z surface, we generate nm-large steps on top of the on-chip WGRs by this final cleaning process. This roughness has a larger effect on the $Q$-factors of the e-polarized modes than on the o-polarized modes, hence the higher losses. This is not a fundamental problem and can be solved by revising of the final polishing and cleaning procedure. 

Due to the advantages regarding $Q$-factor and coupling efficiency, we pump the resonator with o-polarized light to generate frequency-doubled light, which is also o-polarized (type 0 phase matching, ooo). In doing so, we observe the generation of red light in the WGR with a CCD-camera. This light is coupled into the waveguide (Fig.~\ref{fig:Effizienz}). The light at the end facet is collimated and one part is diverted for interrogation by a spectrometer. We use a long-pass filter (cut-on wavelength: 1100 nm), to separate the pump light from the converted one and measure the intensities with a detectors operating in the visible and in the near-infrared spectral regions. The recorded spectrum shows that the wavelength of the converted light is $775.5~\mathrm{nm}$, confirming the second-harmonic generation (SHG). The second-harmonic light is also o-polarized.

Next, we optimized the output power of the converted light by adjusting the distance between the coupling waveguide and the WGR and thus the coupling efficiencies. Furthermore, we optimized the phase matching of the conversion process by varying the resonator temperature. Afterwards, we tuned the pump power from $0.5$ to $5\,\mathrm{mW}$, while recording the power of the converted light, to calculate the conversion efficiency (Fig.~\ref{fig:Effizienz}). The measured efficiency depends linearly on the pump power and the normalized conversion efficiency is $9\times10^{-4}\,\mathrm{mW}^{-1}$. 

\begin{Figure}[h]
	\centering
	\includegraphics[width=8.4 cm]{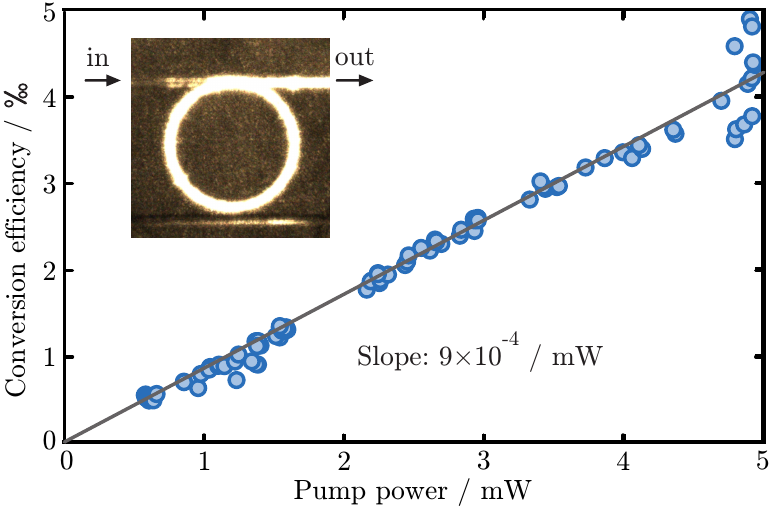}
	\captionof{figure}{Measured conversion efficiency of SHG (blue dots) and theoretical curve fitted by the analytical model. Inset: Image of the WGR while second-harmonic light is generated and coupled out into the coupling waveguide.} 
	\label{fig:Effizienz}
\end{Figure}

\begin{Figure}[htbp]
	\centering
	\includegraphics[width=8.4 cm]{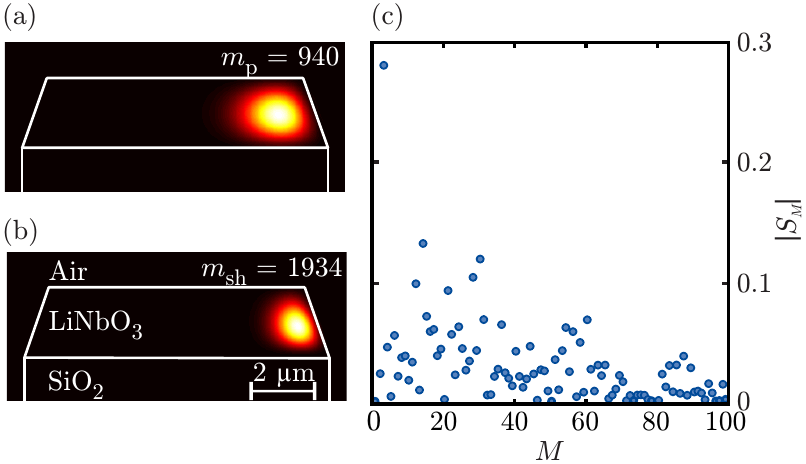}
	\captionof{figure}{Mode cross section of (a) the pump light and (b) the second-harmonic light. (c) Fourier coefficients of a theoretical poling structure with $23\,\mu\mathrm{m}$ periodicity and $5\,\mu\mathrm{m}$ domain width for type~0 phase matching.}
	\label{fig:Fouriekoeffizienten}
\end{Figure}

To compare this result with the theoretically expected value, we start with the relation for the conversion efficiency~$\eta$ \cite{Breunig16}

\begin{equation}
\eta \equiv \frac{P_\mathrm{sh}}{P_\mathrm{p}} = 4\frac{r_\mathrm{sh}}{1+r_\mathrm{sh}} \frac{r_\mathrm{p}}{1+r_\mathrm{p}}\frac{X}{(1+X)^2} \quad\mathrm{with} 
\label{eq:efficiency}
\end{equation}

\begin{equation}
X(1+X)^2 = \frac{r_\mathrm{p}}{1+r_\mathrm{p}}\frac{P_\mathrm{p}}{P_0}.
\label{eq:X}
\end{equation}

where $P_{\mathrm{p}}$ is the pump power, $P_{\mathrm{sh}}$ the power of the second-harmonic light, and $r_\mathrm{p,sh}$ are the ratios between the coupling losses and the internal losses in the resonator. This means that at $r_\mathrm{p,sh} = 1$ we have critical coupling, where the internal loss equals the coupling loss. Furthermore, we have the characteristic power at which high efficiencies for the conversion process can be reached,

\begin{equation}
P_0 = \nu_\mathrm{p}\frac{\pi\epsilon_0 n^4_\mathrm{p} n^2_\mathrm{sh}}{8 d^2_\mathrm{eff}}\frac{1}{Q^2_\mathrm{i,p}Q_\mathrm{i,sh}}V_\mathrm{eff}(1+r_\mathrm{p})^2(1+r_\mathrm{sh}).
\label{eq:P0}
\end{equation}
Here, $\nu_\mathrm{p}$ is the frequency of the pump light, $n_\mathrm{p,sh}$ are the refractive indices of the resonator material, $Q_\mathrm{i,p},Q_\mathrm{i,sh}$ are the unloaded $Q$-factors for the two waves, and $V_\mathrm{eff}$ is the effective mode volume. We have the effective nonlinear coefficient $d_\mathrm{eff} = |S_M|\,d_{222}$ with the nonlinear coefficient of the material $d_{222}=2.1\,\mathrm{pm/V}$ \cite{Nikogosyan05} and a correction constant $|S_M|\leq 2/\pi$, which originates from the quasi-phase matching structure and the crystal symmetry.

In WGRs, the following selection rule for quasi phase matching applies \cite{Breunig16}
\begin{equation}
2m_\mathrm{p}-m_\mathrm{sh}\pm M = 0,
\label{eq:m}
\end{equation}
with the azimuthal mode numbers $m_\mathrm{p}$, $m_\mathrm{sh}$ for the pump and the second-harmonic waves and the integer $M$, provided by the quasi-phase-matching structure. We calculate  $m_\mathrm{p}=940$ and $m_\mathrm{sh}=1934$ (Fig.~\ref{fig:Fouriekoeffizienten}) with a finite-element simulation (\textit{Comsol, Wave Optics Module}). Due to Eq.~(\ref{eq:m}), the linear poling structure has to compensate for $|M| = 54$. We used a poling structure with a periodicity of $23\,\mu \mathrm{m}$ and a domain width of approximately $5\,\mu \mathrm{m}$. Figure~\ref{fig:Fouriekoeffizienten} shows the Fourier coefficients of the poling structure with $|S_{M=54}|=0.06$, thus our structure is indeed expected to enable quasi phase matching. Furthermore, we have an intrinsic $Q$-factor for the pump light of $Q_\mathrm{p} = 1.3\times 10^6$ and since we assume that the $Q$-factor is limited by surface scattering we can estimate the $Q$-factor of the second harmonic light to be $Q_\mathrm{sh}=Q_\mathrm{p}(\lambda_\mathrm{sh}/\lambda_\mathrm{p})^3=Q_\mathrm{p}/8$. We calculate the effective mode volume to be $V_\mathrm{eff} = 3.8\times10^{3}\,\mu\mathrm{m^3}$ with our finite-element simulation. With $r_\mathrm{p} = 1$ and $r_\mathrm{sh}=1$ we can estimate the characteristic power with Eq.~\ref{eq:P0} to be $P_0 = 524\,\mathrm{mW}$, which means that $P_\mathrm{p}\ll P_0$.

From Eq.~(\ref{eq:X}) it follows with $r_\mathrm{p}=1$ that also $X\ll 1$ and hence $X(1+X)^2\approx X$. Taking Eq.~(\ref{eq:X}) and with $P_\mathrm{p}\ll P_0$, Eq.~(\ref{eq:efficiency}) can be approximated to
\begin{equation}
\eta(P_\mathrm{p}) \approx 4\left(\frac{r_\mathrm{p}}{1+r_\mathrm{p}}\right)^2 \frac{r_\mathrm{sh}}{1+r_\mathrm{sh}}\frac{1}{P_0}P_\mathrm{p}. 
\label{eq:efficiencyUndepletedPump}
\end{equation}

A fit to the experimental data gives $r_\mathrm{sh}=0.6$. This is reasonable since we have just one coupling waveguide for both, the pump light and the second-harmonic light and therefore we are able optimize the coupling ratio just for one wave. If we have critical coupling for the pump light, it is obvious that we are in the undercoupled regime for the second-harmonic light due to a faster decay of the evanescent field for light with shorter wavelength and a slightly different mode position in the WGR. Critical coupling for both, the pump light and the second-harmonic light, could be achieved by using a second coupling waveguide, which gives the opportunity to optimize the coupling for the pump and second-harmonic waves independently.

From Eq.~(\ref{eq:efficiencyUndepletedPump}) it becomes evident that a high effective nonlinear-optical coefficient and also high $Q$-factors are crucial for getting large conversion efficiencies. The pump light we use is ordinarily polarized, which means that we employ the nonlinear-optical coefficient $d_{222}$, which is relatively small compared to $d_{333}$. If we can get also high $Q$-factors in the range of one million for modes pumped with e-polarized light, we can expect a very high achievable normalized conversion efficiency of $0.15\,\mathrm{mW}^{-1}$.

In conclusion, we show how to achieve quasi phase matching in on-chip WGRs, to the best of our knowledge, for the first time. We demonstrate SHG with a normalized conversion efficiency of $9\times10^{-4}\,\mathrm{mW}^{-1}$, which is of the same order of magnitude as reported in previous work where cyclic phase matching was used \cite{Lin16}. However, we want to emphasize that compared to cyclic phase matching, we can adjust the period length of the periodical poling structure, which gives us the freedom to tailor phase matching for every other spectral range. Furthermore, it gives full flexibility for the polarization of the interacting light fields so that the largest nonlinear-optical coefficient $d_{333}$ can be accessed or, e.g., polarization entanglement of the interacting waves can be achieved. Thus, the presented procedure for achieving periodically-structured ferroelectric thin films overcomes the phase-matching limitation of on-chip WGRs. This limitation has so far prevented on-chip WGRs from fully unfolding their high potential, which they have due to their cheap and precise manufacturing by standard semiconductor fabrication techniques. Being able to deploy now type~0 and type~II phase matching opens entirely new possibilities for on-chip WGRs and gives also a promising perspective to demonstrate the first on-chip optical parametric oscillator or to realize compact photonic circuits comprising frequency synthesizers.

\section*{Funding Information}
We gratefully acknowledge financial support from the German Federal Ministry of Education and Research (funding program Photonics Research Germany, 13N13648). Richard Wolf appreciates the support by a Gisela and Erwin Sick Fellowship. Yuechen Jia thanks the support from Alexander von Humboldt Foundation.

\end{multicols}
\vspace*{1mm}

\begin{multicols}{2}
	\renewcommand\refname{
\end{multicols}

\newpage

\vspace*{1cm}

{\sffamily\huge\bfseries\noindent Quasi-phase-matched nonlinear optical frequency conversion in on-chip whispering galleries: supplementary material}

	\vspace*{0.3cm}

\textbf{This document provides supplementary information to ?Quasi-phase-matched nonlinear optical frequency conversion in on-chip whispering galleries?.}

\begin{multicols}{2}

\section{Fabrication}
The fabrication of pp-LNoQ substrates is done basically in four steps: domain inversion of a bulk LN chip by field-assisted poling; surface cleaning and functionalizing of the LN chip and of an $\alpha$-quartz substrate; thermal bonding of the LN on quartz; and lapping and polishing of LN to the final thin film.

We write calligraphically a linear grating with $23\,\mu\mathrm{m}$ period length and approximately $5\,\mu\mathrm{m}$ domain width into a 300-$\mu\textrm{m}$-thick 5-mol.$\%$-MgO-doped optically-polished $z$-cut-LN chip (chip size: $17\times18\,$mm$^2$). Here, a 150-nm-thick chromium layer is sputtered on the +$z$-side of the crystal and serves as the backelectrode. We write the domains with a tungsten carbide tip moving along the $y$-crystal axis. While writing the domains, the applied high voltage is controlled in a way that a constant poling current of $30\,\mathrm{nA}$ flows, and the chip temperature is set to $150\,\mathrm{^\circ C}$. After poling the chromium layer is removed by selective wet-chemical etching.

Next we clean the periodically-poled sample and also a z-cut-$\alpha$-quartz chip with acetone and isopropanol, followed by a further cleaning and surface functionalization procedure. It basically consists of three steps: ultra-sonic-assisted rinsing of the samples for 10 minutes in a H$_2$SO$_4$/H$_2$O$_2$ (1:1) solution, then in a HCl/H$_2$O$_2$/H$_2$O (1:1:5) solution and finally in a H$_2$O$_2$/NH$_4$OH/H$_2$O (1:1:5) solution. Each step is followed by rinsing the samples with deionized water for 5 minutes, and we dry the sample with nitrogen in the end.

After bonding of the LN chip on the quartz substrate, we enhance the bonding force by tempering. We heat the sample on a hot plate with $4\,\mathrm{K/min}$ to $325\,\mathrm{^\circ C}$ and temper the sample for 5 hours.

Subsequently, we reduce the thickness of the periodically-poled LN to $2~\mu\textrm{m}$ by lapping and polishing with a standard wafer-polishing machine (\textit{PM5, Logitech}). This is done in three steps. First we use a solution with 9-$\mu\textrm{m}$-Al$_2$O$_3$ particles and a lapping pressure of $1~\textrm{N/cm}^2$ to lap the LN down to a thickness of $12~\mu\textrm{m}$. Next, we reduce the LN-film thickness to approximately $6~\mu\textrm{m}$, deploying a solution with 1-$\mu\textrm{m}$-ceroxide particles and a lapping pressure of $2.5~\mathrm{N/cm}^2$. Last, to achieve substrates with optical grade surface, we use SF1 Syton (\textit{Logitech}) and a polishing pressure of  $2.5~\mathrm{N/cm}^2$ to polish the LN-film down to the final thickness of $2~\mu\mathrm{m}$. The film-thickness measurements are done with a profilometer. The WGRs are structured in the middle area ($6\times8\,$mm$^2$) where we have a film thickness homogeneity of $\pm 250\,\mathrm{nm}$.

\end{multicols}

\end{document}